\documentclass{sig-alternate-05-2015}

\usepackage[
  pass,
]{geometry}

\usepackage[utf8]{inputenc}
\usepackage[T1]{fontenc}

\usepackage{csquotes}
\usepackage[english]{babel}
\usepackage{comment}

\usepackage{microtype}

\usepackage{url}
\usepackage{color}
\usepackage{fixltx2e}
\usepackage[colorlinks,pdfusetitle]{hyperref}
\makeatletter
\hypersetup{pdftitle={\@title},pdfauthor={Flávio Martins, João Magalhães, and Jamie Callan}}
\makeatother

\usepackage{tabularx, booktabs}
\newcolumntype{Y}{>{\arraybackslash}X}
\newcommand{\bftab}[1]{\fontseries{b}\selectfont#1\fontseries{\seriesdefault}}

\usepackage{multirow} 

\usepackage{graphicx,calc}

\usepackage[caption=false]{subfig}

\graphicspath{{figures/}}

\usepackage{float}
\usepackage{placeins}
\usepackage{afterpage}
\usepackage{flushend}

\usepackage{pgfplots}
\pgfplotsset{compat=newest}

\usepackage{cleveref}
\crefname{subsection}{subsection}{subsections}
\crefname{figure}{figure}{figures}
\crefname{equation}{eq.}{eqs.}

\usepackage{amsmath}
\usepackage{amssymb}
\usepackage{mathtools}
\mathtoolsset{centercolon}

\usepackage{mathbbol}

\newcommand*\mean[1]{\bar{#1}}
\newcommand*\rfrac[2]{{}^{#1}\!/_{#2}}

\newcommand{\todo}[1]{}
\renewcommand{\todo}[1]{{\color{red} TODO: {#1}}}

\usepackage[numbers,sectionbib,sort&compress]{natbib}
\makeatletter
\renewcommand\bibsection{%
    \section[References]{
        {References} 
        {\vskip -9pt plus 1pt} 
        \@mkboth{{\refname}}{{\refname}}%
    }%
}%
\makeatother
\newcommand{\superscript}[1]{\ensuremath{^{\textrm{#1}}}}
\def\wu{\superscript{*}}
\def\wg{\superscript{\dag}}
\def\wl{\superscript{*}}
\def\bl{\superscript{\textastdbl}}
\def\wc{\superscript{\dag}}
\def\bc{\superscript{\ddag}}
\DeclareRobustCommand\textastdbl{%
  \leavevmode
  {\sbox0{\ddag}%
   \ooalign{\raisebox{\ht0-\height}{*}\cr
            \raisebox{\depth-\dp0}{\scalebox{1}[-1]{*}}\cr}%
  }%
}
\def\sharedaffiliation{%
\end{tabular}
\begin{tabular}{c}}
\CopyrightYear{2016} \setcopyright{acmcopyright}
\conferenceinfo{WSDM'16,}{February 22--25, 2016, San Francisco, CA, USA.}
\isbn{978-1-4503-3716-8/16/02}\acmPrice{\$15.00}
\doi{http://dx.doi.org/10.1145/2835776.2835825} 

\begin{document}







\title{Barbara Made the News: Mining the Behavior of Crowds for Time-Aware Learning to Rank
    \thanks{Please cite the WSDM 2016 version of this paper.}
}

\numberofauthors{3}
    \author{
        \alignauthor \ Flávio Martins\wu\\
        \email{flaviomartins@acm.org}
        \alignauthor \ João Magalhães\wu\\
        \email{jm.magalhaes@fct.unl.pt}
        \alignauthor \ Jamie Callan\wg\\
        \email{callan@cs.cmu.edu}
        \sharedaffiliation
        \begin{tabular}{ccc}
            \affaddr{{\wu}NOVA LINCS, Department of Computer Science{\ }}           & & \affaddr{{\wg}Language Technologies Institute{\ }} \\
            \affaddr{Faculty of Science and Technology} & & \affaddr{Carnegie Mellon University} \\
            \affaddr{Universidade NOVA de Lisboa}       & & \affaddr{5000 Forbes Avenue} \\
            \affaddr{2829-516 Caparica, Portugal}       & & \affaddr{Pittsburgh, PA 15213, USA} \\
        \end{tabular}
    }
\maketitle
\begin{abstract}
In Twitter, and other microblogging services, the generation of new content by the crowd is often biased towards immediacy: \emph{what is happening now}.
Prompted by the propagation of commentary and information through multiple mediums, users on the Web interact with and produce new posts about newsworthy topics and give rise to trending topics. 
This paper proposes to leverage on the behavioral dynamics of users to estimate the most relevant time periods for a topic.
Our hypothesis stems from the fact that when a real-world event occurs it usually has \emph{peak times} on the Web: a higher volume of \emph{tweets}, new visits and edits to related Wikipedia articles, and news published about the event.

In this paper, we propose a novel time-aware ranking model that leverages on multiple sources of crowd signals.
Our approach builds on two major novelties. 
First, a unifying approach that given query $q$, mines and represents temporal evidence from multiple sources of crowd signals. This allows us to predict the temporal relevance of documents for query $q$.
Second, a principled retrieval model that integrates temporal signals in a learning to rank framework, to rank results according to the predicted temporal relevance.
Evaluation on the TREC 2013 and 2014 Microblog track datasets demonstrates that the proposed model achieves a relative improvement of 13.2\% over lexical retrieval models and 6.2\% over a learning to rank baseline.
\end{abstract}



\vfill\eject
\section{Introduction}
\label{sec:introduction}

A networked world and the pervasiveness of Internet access enables the rapid adoption of new online communication mediums.
People are increasingly sharing new information via microblogging services and other similar social media services.
Events are discussed in real-time, thus search became a pressing demand from Web users in general.


Standard search techniques, such as retrieval schemes based on language modeling have proven to be very effective in Web documents.
However, standard text retrieval functions have under-performed in the ranking of microblog posts, because \emph{tweets} are short and highly biased by real-world events.
Previous research in time-aware ranking explored the assumption that more recent documents are more relevant \cite{li_time_2003}. 
Later models revised this assumption in line with what is observed in Twitter and other real-time document collections: for time-sensitive queries, documents tend to cluster temporally~\cite{dakka_answering_2012, efron_temporal_2014}.
Most prior work gather temporal information from a single source, typically either the corpus itself (e.g., Twitter) or Wikipedia (e.g., in the form of edits or views). These strategies assume that all important events will have an impact on Twitter or Wikipedia, and that the temporal signal available from Twitter or Wikipedia will be clear and unambiguous, however that is not always the case. 

We depart from previous approaches, by observing that events have an impact not only on Twitter but also in other Web domains such as news, clicks, query logs, etc.
Our approach is based on the intuition that discussions about an event (and therefore relevant documents) are more likely to occur around the same time periods across multiple social-media services.
Thus, it is reasonable to assume that it would be advantageous to expand these methods to further integrate external sources, to offer more context to the \emph{tweets} as well as to the users' queries intent.
Temporal queries may not exhibit temporal patterns of user search behavior in query logs or even in the temporal density of an initial set of retrieved documents, but still contain underlying temporal information needs.
In this context, we aim at refining the temporal relevance density function of a time-sensitive query by using multiple external Web sources.

This approach brings a series of novel contributions: (1) the mining of crowd signals from different Web sources (e.g., news, Wikipedia articles, and Twitter) to detect relevant time periods, and (2) a unifying method to generate temporal features that can be used for multiple time-aware tasks, and (3) a time-aware ranking model based on learning to rank. 
Evaluation on the TREC 2013 and TREC 2014 Microblog Track datasets has shown that the proposed retrieval model outperforms state-of-the-art methods. More specifically, we found that most queries benefited from at least a second source of temporal information.

This paper is organized as follows: in section 2 we briefly analyze an event that motivates the proposed method; in section 3 we present the related work; section 4 details the mining of temporal crowd signals; the ranking framework is formalized in section 5 where we detail ranking with multiple temporal features; evaluation is presented in section 6; and a more fine-grained discussion of results in section 7.

\section{Barbara Made the News}
\label{sec:barbara}
\begin{figure}[!t]
	\centering
	\subfloat[Temporal evidence from multiple Web sources.]{
		\begin{minipage}{\columnwidth}
			\centering
			\includegraphics[trim=5 5 0 5, clip=true, width=0.95\textwidth]{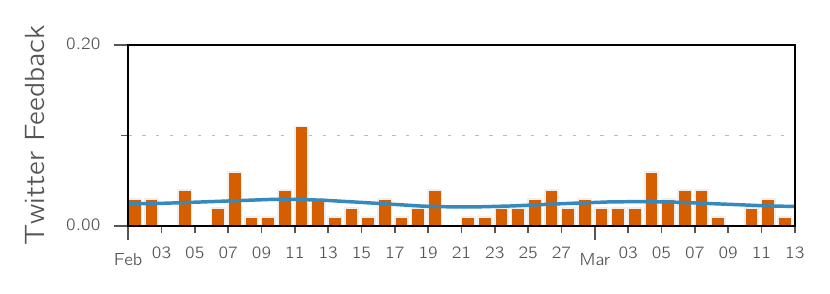} \\
			\includegraphics[trim=5 5 0 5, clip=true, width=0.95\textwidth]{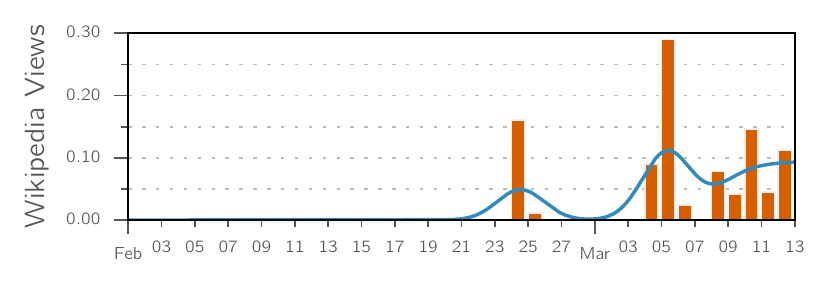} \\
			\includegraphics[trim=5 0 0 0, clip=true, width=0.95\textwidth]{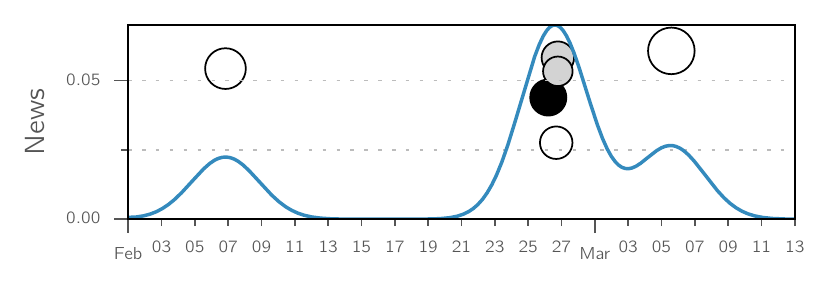} \\
			\label{fig:temporal-evidence-deconstructed-topic}
		\end{minipage}
	}
	
	\subfloat[Temporal distribution of relevant tweets (qrels).]{
		\begin{minipage}{\columnwidth}
			\centering
			\includegraphics[trim=5 5 0 5, clip=true, width=0.95\textwidth]{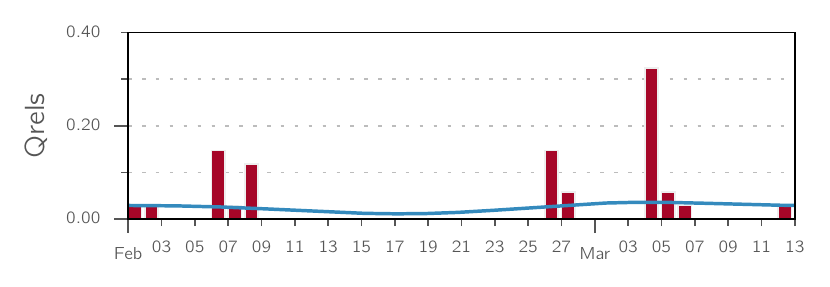} \\
			\label{fig:relevant-deconstructed-topic}
		\end{minipage}
	}
	\caption{Results for query \enquote{Barbara Walters, chicken pox} on different sources. The histograms indicate the volume of data for each source per day. In the News source, each circle corresponds to a matching news article, its size is proportional to the Jaccard index between the query and the news headline. Its color indicates the news site (Reuters -- gray, Associated Press -- black and USA Today -- white). The curves depict a smoothed estimate.
	}
	\label{fig:deconstructed-topic}
\end{figure}

On February~2013 it was revealed that Barbara Walters, a well-known figure in the U.S. television, got chicken pox.
This real-world event sparked multiple processes on the Web, such as the propagation of news articles about the event, commentary on Twitter and an increased interest in the article page for \enquote{Barbara Walters} on Wikipedia.
Often, these signals can be mined as the event unfolds and for some Web sources can also be mined retrospectively.  
In \Cref{fig:deconstructed-topic} we dissect how multiple Web sources can show signals related to the crowd behavior in reaction to this newsworthy event.

When searching Twitter for \enquote{Barbara Walters, chicken pox}, we cannot make out from the visualization of \emph{Twitter} (the distribution of initially retrieved documents) any of the time periods that are actually relevant (the distribution of relevant tweets).
Having found no useful temporal evidence using only the collection (i.e., Twitter Feedback) we turn our attention to extra external sources that can help disambiguate the days that are temporally relevant to this query.

The \emph{Wikipedia views} signal chart for this topic shows the volume of page views per-day of the \enquote{Barbara Walters} article on Wikipedia.
We see that the biggest spike was dominated by March~4, and in the following days the interest in the page decreases. This actually corresponds to the highest volume day in the ground-truth, albeit with about a day of lag.

Turning our attention to \emph{News} sources we are able to spot three (3) main time periods that are good candidates to better model temporal relevance.
The first peak corresponds to a news article with alarming news about \emph{chicken pox} mentioning \emph{Barbara Walters}, which accompanied the discussion of the topic on Twitter, the second peak corresponds to the propagation of news that Barbara Walters is recovering and the third peak is related to the news of her return to the U.S. television talk show \enquote{The View} on ABC.

To verify the relevance of each time period evidenced by each source, we can examine the volume of relevant tweets per day on Twitter using the list of relevant documents (Qrels), \Cref{fig:relevant-deconstructed-topic}. The key insight from this comparison is that two temporal sources have useful information providing different, but relevant information.
This analysis hints that mining multiple sources of evidence to detect when events are \emph{happening} can be more robust than using only a single source of evidence (i.e., the corpus).

The rationale supporting our research hypothesis calls for a method that overcomes this problem by disambiguating the relevant time periods using multiple sources of evidence: internal and external, to improve retrieval in time-sensitive scenarios such as search on Twitter.
Our approach is to mine useful temporal features to propose a novel time-aware ranking method with multiple temporal sources (RMTS).

\section{Background and Related Work}
\label{sec:relatedwork}

One of the main issues in microblogs is the quality of the posts themselves.
Removing duplicates (i.e. \emph{retweets}), \emph{tweets} written in unwanted target languages, and spam improves the ranking performance \cite{kim_overcoming_2012}.
Some techniques~\cite{massoudi_incorporating_2011} estimate the \emph{tweet's} quality using textual features such as the presence of emoticons, slang and unknown vocabulary, following prior research on blog post retrieval \cite{weerkamp_credibility_2012}.
More elaborate approaches~\cite{weng_twitterrank_2010, efron_information_2011} measure the influence of authors instead, using features such as the number of followers and the number of \emph{retweets} to estimate the author's credibility.

Another main issue is the temporal relevance of the microblog posts.
Previous research \citep{teevan_twittersearch_2011} has revealed that people search Twitter to find temporally relevant information, monitor popular trends, discover breaking news and follow events. 
Therefore, it is imperative that the design of search tools for social media incorporates the time dimension.

Existing works on time-aware ranking can be classified according to two main notions of relevance with respect to time \cite{nattiyakanhabua_temporal_2015}: 
1) recency-based ranking, and
2) time-dependent ranking. 
Recency-based ranking usually improves scoring in social media and microblogs.
Most topics searched by users on Twitter seem to imply a real-time information need \cite{teevan_twittersearch_2011}, therefore recency ranking methods can be useful in social media search.
Recency can be easily integrated into text retrieval with the time-based language model method proposed by~\citet{li_time_2003}, an extension to the standard query-likelihood model~\cite{ponte_language_1998, zhai_study_2004}.
The use of \emph{recency priors} to promote recently published documents proposed by \citet{li_time_2003} assumes that in production systems, documents published closer to the time of the query are more relevant to the query.
Given the document $D$ timestamp $T_D$, they propose to model $P(D)$ in the query-likelihood model via the exponential distribution, $P(D) = \lambda e ^{-\lambda (T_C-T_D)}$, where $\lambda \geq 0$ is the decay rate parameter of the exponential distribution and $T_C$ is the most recent date in the collection.
In the microblog search scenario \citet{efron_estimation_2011} found that $\lambda = 0.01$ provided the best average performance.
This is a simple approach that usually outperforms the standard query-likelihood model.

However, \citet{jones_temporal_2007} notes that queries that favor recency are just a subset of the time-sensitive queries.
\citet{dakka_answering_2012} identified the need to find the important time periods for time-sensitive queries and to integrate this temporal relevance in the ranking model. 
Their ranking model explicitly divides documents into its lexical and temporal evidences: $W_D$, the words in the document and $T_D$, the document's timestamp, leading to $P(D|Q) \sim P(W_D|Q) P(T_D|Q) P(W_D) P(T_D)$.
\citet{dakka_answering_2012} proposes techniques to estimate the $P(T_D|Q)$ from histograms, however the calculation of the histogram bins is linked to many parameters.


Additionally, a few time-based pseudo-relevance feedback methods were proposed for retrieval in time-sensitive collections.
\citet{whiting_temporal_2012} combines pseudo-relevant document term distribution and temporal collection evidence using a variant of PageRank over a weighted graph that models the temporal correlation between n-grams.
Another approach by \citet{peetz_using_2013} leverages on the temporal distribution of the pseudo-relevant documents themselves.
For each query, bursty time periods are identified and then documents from these periods are selected for feedback. 
The query model is updated with new terms sampled from the higher quality documents selected.

\citet{efron_temporal_2014} proposed a general and principled retrieval model for microblog search with temporal feedback. Their approach models the temporal density of a query $f(t|q)$ with a kernel density estimate, with all the advantages brought by this method: the natural smoothness of the resulting function and a fully automated way to estimate the model variables (bandwidth selection is data-driven, i.e: a function of the initial rank). This estimated temporal relevance is then employed to rerank documents with a log-linear model. 

In contrast to previous work, we propose to use multiple external Web sources to robustly identify the relevant time periods for each query, instead of relying only on the temporal distribution of pseudo-relevant documents.
Our method also accounts for the underestimation of the density at the boundaries, which can negatively affect queries with relevant documents near temporal extremes (e.g., query-time).

There are several previous works employing feature-based methods and machine learning for ranking with temporal features. \citet{dai_learning_2011} propose to run each query against a set
of rankers, which are weighted based on the temporal profile of a query, and therefore minimize the risk of degraded performance due to misclassifying the query in terms of recency intent.
Recent time-dependent ranking approaches have resorted to learning to rank techniques that exploit non-temporal and temporal features~\cite{kanhabua_learning_2012, costa_learning_2014}.

\section{Temporal Crowd Signals}
\label{sec:crowds}
The starting point of our hypothesis is that many real-world events originate a burst of simultaneous Web activity in multiple, possibly heterogeneous, data sources. 
The analysis of multiple temporal evidences is a powerful tool that can allow us to stitch together information from these sources \cite{bendersky_effective_2012}. The first step towards a solution requires the design of a unified representation of temporal signals extracted from several different sources.

\subsection{A Unified Representation of \\Temporal Crowd Signals}
Considering a set $S \in \{s_1, s_2, \ldots, s_k, \ldots\}$ of information sources that reflect a real-world event, our goal is to discover the relevance of a timestamp for a particular query. In other words, we wish to infer a function
\begin{equation}
f_{s_k}(q_a, t_b) \in [0,1]
\end{equation}
for each source $s_k$, to compute a relevance correlation score for the query $q_a$ at instant $t_b$ (this will be formalized as document $d_b$ timestamp).
To devise the $f_{s_k}$ functions, we need to mine the behavior of the crowds on each source to discover signals that are related to the query at hand. In our framework, a signal corresponds to any user activity that is registered by the information source, e.g., a Wikipedia page view, a tweet, or the publication of a news article. Once we discover these signals, their timestamps and relevance to the query are collected into two sets:
\begin{equation}
T^{s_k} = \{t_1^{s_k},t_2^{s_k},\ldots,t_n^{s_k}\} \hspace{3mm}
W^{s_k} = \{w_1^{s_k},w_2^{s_k},\ldots,w_n^{s_k}\}
\end{equation}
Each pair $(t_i^{s_k}, w_i^{s_k})$ fully characterizes a temporal signal -- it saves the signal $i$ timestamp and its importance to the query that originated the mining process.

For each source of temporal signals $s_k$ and its set of (mined) timestamps $T^{s_k} = \{t_1^{s_k},t_2^{s_k},\ldots,t_n^{s_k}\}$ we propose to use the same unified representation.
For each document $d_b$ we would like to find $P(r|t_{d_b},q_a,s_k)$, the probability of relevance of its timestamp $t_{d_b}$ according to source $s_k$ and given query $q_a$.
The estimation of the joint distribution $f_{s_k}(q_a,t)$ over the time span of the corpus is key.
The probability density function $f_{s_k}(t_{d_b}|q_a)$ is approximated by a kernel density estimation which is advantageous due to the natural smoothness of the resulting function:
\begin{equation}
\hat{f_{s_k}}(t) = \frac{1}{nh} \sum_{i=0}^n w_i^{s_k} K \left( \frac{t - t_i^{s_k}}{h} \right)
\end{equation}
where $t_i^{s_k}$ are the timestamps mined from a source $s_k$, the kernel function $K ( z )$ corresponds to the Gaussian kernel $\mathcal{N} \left( z, 0 \right)$, and the optimal bandwidth $h$ can be estimated by a data-driven method such as Silverman's rule-of-thumb $h^* \approxeq 1.06\, \hat{\sigma}\, n^{-\rfrac{1}{5}}$.
Finally, the vector $\{w_1^{s_k},w_2^{s_k},\ldots,w_n^{s_k}\}$, of non-negative weights on the $t_i^{s_k}$'s timestamps, weights each timestamp by its importance (e.g., to a query).

\subsection{Temporal Signals from Multiple Sources}
Most microblog queries are temporally spiky, following live events related to breaking news, celebrities, other entities and events, periodic queries, e.g., TV shows, and ongoing events.
Our approach is volume-based and we assume as a starting point that topics bursting on Twitter are correlated with News and a higher volume of page views and edits for relevant Wikipedia pages.

To extract temporal crowd signals that are relevant to the search query, one needs to design source-specific methods that filter candidate signals and compute their relevance to the search query. This is the process that we discuss next.

\subsubsection{News}
Our system indexes the news headlines produced by multiple news sources incrementally.
We select a small number of high signal-to-noise ratio sources of News: the Associated Press (AP), BBC's UK and World (BBC), Reuters and USA Today.
We used an automatic rule-based extraction method to extract explicit non-relative temporal expressions \cite{schilder_temporal_2003} to detect the publication date of the news articles.
A per-site timezone was set manually to adjust the site's time offsets. 

For a given query $q_a$, we focus on finding a set of news headlines matching the search query, therefore we first retrieve a set of candidate headlines $H = \{h_1,h_2,\ldots,h_n\}$ from our index using a standard retrieval method.
The temporal signals of the news source $s_h$ are represented by the set $T^{s_h} = \{t_1^{s_h}, t_2^{s_h},\ldots, t_n^{s_h}\}$ of headline timestamps.

The timestamp of each headline $t_i^{s_h}$ should be weighted according to its relevance to the query.
Therefore, we measure the similarity of each headline to the query using the Jaccard similarity coefficient because usually in both headlines and queries $TF_i = 1$.
This choice also avoids overweighting headlines with repeated words.
The Jaccard index of Jaccard similarity coefficient of two sets of words $A$ and $B$ is given by $J(A, B) = \frac{|A \cap B|}{ |A\cup B|}$.
Thus, the weight $w_i^{s_h}$ of each timestamp $t_i^{s_h}$ is computed as the Jaccard similarity coefficient $J(q_a, h_i)$.

\subsubsection{Wikipedia}
To find a Wikipedia article relevant to the query $q_a$, our system searches for article titles on Wikipedia using their API.
The first 10 candidate page titles returned are ranked using the Jaccard similarity coefficient and the most similar page title is selected.
We hypothesize that users searching on Twitter and on the Web using text queries do so using similar queries and we try to model this behavior.
Following this rationale we posit that, especially in the case of page views, if we don't find the article that best matches the query we should still get some feedback from alternative articles, because users are also likely to have interacted with them when searching the Web for information about a topic that is trending on Twitter.

\textbf{Wikipedia page views.}
\label{sec:wikipediaviews}
To obtain page view statistics of Wikipedia articles we use a public API\footnote{\url{http://stats.grok.se/}} that provides histograms of counts binned by day (we will refer to this source as $s_v$). 
For our temporal signal extraction we fetch data from 1~December 2012 up until the time of the query, therefore the temporal signals for a query can go up to 4 months of data.
To eliminate noise we normalize the frequencies using the mean of article view counts per day and obtain a normalized set of view counts per-day $W^{s_v} = \{w_1^{s_v},w_2^{s_v},\ldots,w_n^{s_v}\}$.
Finally, the temporal signals extracted from page views are represented as $T^{s_v} = \{t_1^{s_v}, t_2^{s_v},\ldots, t_n^{s_v}\}$.
In this case there is one timestamp per-day over the past months. 
To minimize error, we set the timestamp to midnight.

\textbf{Wikipedia page edits.}
\label{sec:wikipediaedits}
In the case of Wikipedia page edits, source $s_e$, we fetch the latest revisions before the time of the query of the Wikipedia article selected, via the Special:Export\footnote{\url{https://en.wikipedia.org/wiki/Special:Export}} feature on Wikipedia.
We process the XML response and calculate the changes between each consecutive pairs of revisions in sequence and filter out article changes signed-off by known automatic Wikipedia Bots\footnote{\url{https://en.wikipedia.org/wiki/Wikipedia:Bots}}.

For the purposes of processing this temporal signal we use $TF(q,e)$, the term-frequency of terms in the query matching the text added in an edit $e$.
This approach aims to restrict the edits, and hence the timestamps to the ones we find a stronger evidence of relatedness.
Following this intuition, we treat terms in the Wikipedia article title as stop words for the calculation of $TF(q,e)$ (e.g., for page \enquote{Barbara Walters} changes matching only \enquote{Barbara} and/or \enquote{Walters} are not counted).
The temporal signals extracted from page edits is given by $T^{s_e} = \{t_1^{s_e}, t_2^{s_e},\ldots, t_n^{s_e}\}$.
The weight $w_i^{s_e}$ of each timestamp is computed as the $\sum_j TF_j(q_a,e)$, the number of terms shared by the query and the edit.

\subsubsection{Twitter}
\label{sec:twitterfeedback}
The crowd signals from Twitter, source $s_t$, are essential to assess the time periods that are more relevant to a query. 
Recent works on time-aware ranking for \emph{tweet} search \cite{dakka_answering_2012,efron_temporal_2014} look at the ranked list of Tweets produced by a standard retrieval model and their timestamps for temporal feedback.
These methods are rooted in the assumption that a search query will originate two distinct distributions, a lexical one and temporal one, that must be integrated into a single rank. 
The way we mine temporal crowd signals from Twitter is inspired by this family of approaches. The set of temporal signals are extracted from tweets retrieved with query $q_a$ -- the temporal signals $T^{s_t} = \{t_1^{s_t}, t_2^{s_t},\ldots, t_n^{s_t}\}$ collect the tweets' timestamps.
The weight $w_i^{s_t}$ of each timestamp is obtained according to the score computed with the query-likelihood retrieval model, i.e.
$w_i^{s_t} = P(q_a|d_b) / \sum_{j=1}^n P(q_a|d_j)$

\section{Ranking Framework}
\label{sec:approach}
Our search model follows a learning to rank (LTR) framework that integrates text features, domain specific features (e.g. number of hashtags) and temporal features extracted from crowd behaviors. 
In this section, we will first discuss the learning to rank model, followed by the description of a comprehensive set of non-temporal features, and will then address the computation of the model parameters.

\subsection{Ranking with Multiple Temporal Signals}
Typical LTR models consider a diverse set of features from the corpus. In this paper we integrate the features into a linear model, where the retrieval score of document $d_b$ for a given query $q_a$ is given by
\begin{align} \label{eq:rs}
\begin{split}
LTR(q_a,d_b) = \sum_i \alpha_i f_{l_i}(q_a,d_b) + \sum_j \beta_j f_{c_j}(d_b)
\end{split}
\end{align}
The set of lexical features,  $f_{l_i}$ covers several text statistics and retrieval scores. Learning to rank literature has also extensively shown that non-lexical features $f_{c_j}$, such as clicks and number followers, are essential to capture the relevance of a document.

In the previous section, we saw how temporal evidence can be mined from data generated by the crowds. \Cref{tab:temporal-features} summarizes the set of temporal feature signals. We proposed a unified view of how temporal signals from the crowds can be represented in a common way. Following this reasoning, we are now ready to plug-in multiple temporal evidence into the initial retrieval model, by extending it to support the time dimension:
\begin{align} \label{eq:rmts}
\begin{split}
RMTS(q_a,t_{d_b},d_b) &= \sum_i \alpha_i f_{l_i}(q_a,d_b)
                  + \sum_j \beta_j f_{c_j}(d_b) \\
                  &+ \sum_k \gamma_k f_{s_k}(q_a,t_{d_b})
\end{split}
\end{align}
where $t_{d_b}$ is the timestamp of document $d_b$ , $f_{s_k}$ returns the likelihood that day $t_{d_b}$ is relevant to the query $q_a$ under the distribution of the crowd source $s_k$. The crowd source $s_k$ indexes one of the temporal sources described in the previous section. The coefficients $\alpha_i$, $\beta_j$ and $\gamma_k$ correspond to the feature weights. 

The RMTS model (Ranking with Multiple Temporal Signals) is composed of three independent parts that capture different statistics of the information domain. 
This model is a well grounded method that generalizes the integration of multiple temporal evidences into a single unified retrieval model.
At this point, it becomes clear the divide between previous work, relying entirely in the corpus timestamps, and the proposed approach that looks at the crowd behaviors outside the corpus to better estimate the temporal relevance of the query. The factorization of crowd temporal information into a rich set of temporal signals, provides a more robust way to disambiguate the relevant days for each search query.

\subsection{Non-temporal features} 
The non-temporal features in \Cref{tab:non-temporal-features} include text-similarity scores computed over the tweet-text corpus and domain specific features such as the number of hastags in the tweet-text or the number of followers of the tweet's author.

The first set of features consider popular retrieval baselines that have proven to be effective in many scenarios, including temporal search with specific additions to address the temporal dimension.
BM25 is our first text retrieval features, one of the most popular text retrieval models.
In the language modeling approach \cite{ponte_language_1998, zhai_study_2004} to retrieval, a language model is created for each document and the ranking is based on the probability of generating the query. The language model with Dirichlet smoothing \cite{zhai_study_2004} is the third text feature baseline - it is also the baseline model in TREC Microblog.

In the Twitter domain, we found that rare words are actually critical for particular queries.
To address this phenomena, we isolated the Inverse Document Frequency (IDF) as a separate feature and not just embedded in other models such as BM25. This way, we allow the LTR framework to best weight the importance of rare words. The length of the tweet is the last text-based feature.

Besides text-based features, \Cref{tab:non-temporal-features} also includes a commonly used microblog-specific features computed over the contents of the tweets' text and users' metrics. This set of features captures the number of mentions, number of followers, number of URLs among other microblog-specific features.

\begin{table}[!t]
	\centering
	\begin{tabular}{ll}
		\toprule
		Feature name & Feature description \\
		\midrule
		BM25 & Okapi BM25 score for tweet-text. \\
		LM.Dir & Language modeling score for tweet-text. \\
		IDF & Sum of term IDF in tweet-text. \\
		Length & Tweet-text length. \\
		\midrule
		NumURLs & Number of URLs in tweet-text. \\
		HasURLs & True if tweet-text contains URLs. \\
		NumHashtags & Number of Hashtags in tweet-text. \\
		HasHashtags & True if tweet-text contains Hashtags. \\
		NumMentions & Number of Mentions in tweet-text. \\
		HasMentions & True if tweet-text contains Mentions. \\
		isReply & True if tweet-text is a reply. \\
		\midrule
		NumStatuses & Number of user's statuses. \\
		NumFollowers & Number of user's followers. \\
		\bottomrule
	\end{tabular}
	\caption{Non-temporal ranking features used by Coordinate Ascent in the RMTS method and other methods based on learning to rank.}
	\label{tab:non-temporal-features}
\end{table}

\begin{table}[!t]
	\centering
	\begin{tabular}{ll}
		\toprule
		Feature name & Feature description \\
		\midrule
		Recency (\textbf{R})          & Recency prior \cite{li_time_2003}. \\
		\midrule
		Twitter Feedback (\textbf{TF}) & Temporal feedback \cite{efron_temporal_2014}. \\
		Wikipedia Views (\textbf{WV}) & Wikipedia article page views. \\
		Wikipedia Edits (\textbf{WE}) & Wikipedia article page edits. \\
		News                          & News headlines. \\
		\bottomrule
	\end{tabular}
	\caption{Temporal ranking features used by Coordinate Ascent in the RMTS methods. All the features (except for Recency) are produced using kernel density estimation of time series extracted from: the initially retrieved timeline, Wikipedia and News.}
	\label{tab:temporal-features}
\end{table}

\subsection{Computing the Model Coefficients}
First, we turn our attention to the problem of correctly estimating the density of individual temporal sources. The temporal histograms of the different sources are quite heterogeneous, indicating different trends and densities across the timeline. Thus, it also required us to revise the standard KDE method and include boundary correction methods~\cite{cwik_data_1993}.

Second, we address the computation of the model coefficients ($\alpha_i$, $\beta_j$ and $\gamma_k$) weighting the contribution of its corresponding feature (temporal or non-temporal) to the final score. To estimate the values of the coefficients we first observed how these sources of temporal signals are related. A real-world event initiates a series of social-media content that users exchange and access sequentially, for example, a news article may trigger a series of Wikipedia page views and the posting of several \emph{tweets}. This cascade effect suggests that temporal sources might cover the event timeline sequentially. Following this reasoning, we computed the coefficients with coordinate ascent to optimize the MAP retrieval measure.

Coordinate ascent~\cite{metzler_linear_2007} is a learning to rank method that is often used to optimize a retrieval metric directly. A recent work uses it to rank microblog posts using quality features~\cite{choi_quality_2012} and it has been shown to obtain higher performance compared to other learning to rank methods on Twitter datasets~\cite{xu_hltcoe_2014}.
On each iteration, the coordinate ascent algorithm moves towards the optimal solution along one coordinate at a time. In other words, each coefficient is updated individually while all the others are fixed. This process is repeated until convergence and using several random starts to avoid local minima.

\subsection{Query dependent rankers}
In the current formalization, we assume that all external sources are equally relevant to every query, i.e. the $\gamma_k$ coefficient is constant for every query.
A deeper inspection of the queries quickly reveals that some queries are more temporal than others. Thus, based on the crowd temporal signals, we grouped queries into temporal vs non-temporal and trained two different LTR models for each group of queries. 
This procedure further expands the temporal reasoning towards the query analsys. 
Our assumption is that a query is classified as being temporal if the news sources returns news matching the query. Otherwise, the query is classified as being atemporal. This distinction enables us to train two models: the first model with temporal features and a second model without such features.
Now, at query time, the system can decide to use the RMTS model (\Cref{eq:rmts}) or the LTR model (\Cref{eq:rs}) that has no temporal features.

\begin{table*}[t]
	\centering
	\begin{tabularx}{0.85\textwidth}{l *{8}{Y}}
		\toprule
		& \multicolumn{2}{c}{Temporal}                      & \multicolumn{2}{c}{Atemporal}                     & \multicolumn{2}{c}{T+A}                           & \multicolumn{2}{c}{All}                           \\ \midrule
		Method & \multicolumn{1}{c}{MAP} & \multicolumn{1}{c}{P30} & \multicolumn{1}{c}{MAP} & \multicolumn{1}{c}{P30} & \multicolumn{1}{c}{MAP} & \multicolumn{1}{c}{P30} & \multicolumn{1}{c}{MAP} & \multicolumn{1}{c}{P30} \\ \midrule
		\multicolumn{9}{c}{Text retrieval baselines} \\ \midrule
    	BM25~\cite{robertson_okapi_1994}      & 0.4054 & 0.6202 & 0.4319 & 0.6392 & 0.4136 & 0.6261 & 0.4136 & 0.6261 \\
        IDF                                   & 0.4275 & 0.6561 & 0.4360 & 0.6235 & 0.4301 & 0.6461 & 0.4301 & 0.6461 \\
        LM.Dir~\cite{zhai_study_2004}         & 0.4331 & 0.6491 & 0.4112 & 0.6020 & 0.4264 & 0.6345 & 0.4264 & 0.6345 \\ \midrule
        \multicolumn{9}{c}{Temporal ranking baselines} \\ \midrule
        Recency~\cite{li_time_2003} & 0.4429  & 0.6667 & 0.4152 & 0.6196 & 0.4343 & 0.6521 & 0.4297 & 0.6552 \\
        KDE(score)~\cite{efron_temporal_2014} & 0.4621 & 0.6711 & 0.4030 & 0.5961 & 0.4438 & 0.6479 & 0.4455 & 0.6509 \\ \midrule
        \multicolumn{9}{c}{Learning to rank models} \\ \midrule
        LTR (\Cref{eq:rs})                    & 0.4688 & 0.6991 & 0.4308 & 0.6216 & 0.4571 & 0.6751 & 0.4528 & 0.6703 \\
        RMTS (\Cref{eq:rmts})                 & 0.5011\bc\bl & 0.7254\bc & 0.4422 & 0.6353 & \bftab{0.4829}\bc\bl & 0.6976\bc\wl & \bftab{0.4809}\bc\bl& 0.6939\bc\wl \\ \bottomrule
    \end{tabularx}
	\caption{Temporal ranking methods results. Symbols \textnormal{\wc} and \textnormal{\wl} stand for a $p < 0.05$ statistical significant improvement over KDE(score) and LTR respectively (\textnormal{\bc} and \textnormal{\bl} for $p < 0.01$). }
	\label{tab:temprank}
\end{table*}

\section{Evaluation}
\label{sec:evaluation}
This section presents the evaluation of the methods described in the previous sections on the TREC microblog search test-bed.
In the TREC microblog ad-hoc search task, the user wishes to find the most recent and relevant posts.
The task can be summarized as: at time $t$, find \emph{tweets} about topic $q$.
Therefore, systems should favor highly informative \emph{tweets} relevant to the query topic that were published before the query time.
Our experiments delve into the problem of reranking \emph{tweets} sampled using a standard retrieval method i.e., query-likelihood taking into account temporal crowd signals from different sources.

\subsection{Datasets and Protocol}

\textbf{TREC datasets.} The experiments were performed with the Tweets2013 corpus using the query topics of the TREC 2013 and TREC 2014 editions of the Microblog track.
The Tweets2013 corpus was created for the TREC 2013 edition and is much larger (240 million \emph{tweets}) than Tweets2011 (16 million \emph{tweets}) used in TREC 2011 and TREC 2012.
The corpus was created by crawling the public stream sample via the Twitter streaming API over a period spanning from 1~February, 2013 - 31~March, 2013 (inclusive).
The TREC Microblog queries are mainly queries about entities and events.
NIST provided relevance judgments over 60 topics for TREC 2013 and 55 topics for TREC 2014 on a three-point scale of \enquote{informativeness}: not relevant, relevant and highly relevant.

\textbf{Filtering Duplicates and Languages.}
In the test collection used, \emph{retweets} are considered not relevant because they are seen as duplicate documents.
Therefore, we filtered \emph{Twitter-style} \emph{retweets} using the tweet metadata available and we also filter out \emph{RT-style} \emph{retweets} that start with \emph{RT}.

Assessors evaluated only relevant \emph{tweets} written in English, therefore we use a language filter to remove \emph{tweets} in other languages.
To build this filter we used the language detection library \mbox{ldig}\footnote{\url{http://github.com/shuyo/ldig}} with a trained model for 19 languages.

\textbf{Sources.} The sources of temporal evidence are the corpus (Twitter), Wikipedia page views and edits, BBC, USA Today, CNN and Associated Press news articles.

\textbf{Protocol.} To allow the comparability to previous and future work with the same datasets we do training on the TREC 2013 topics and test on the TREC 2014 topics.
For learning to rank we used Coordinate Ascent set to optimize mean average precision (MAP) using 20\% of the training data for validation.
In our experiments we follow TREC and report the MAP and P30 results.
Statistical significance of effectiveness differences are determined using two-sided paired \textit{t}-tests following the recommendations by \citet{sakai_statistical_2014}.

\subsection{Baselines}
The first baseline is the \textbf{BM25}~\cite{robertson_okapi_1994} retrieval function parametrized with $k_1=1.2$ and $b=0.75$.
The second, is the \textbf{IDF} retrieval function isolated, which has advantages for \emph{tweets} retrieval.
It favors rare words, a common trait in microblogs, was also included in the experiments as a second baseline.
The third baseline is the query likelihood retrieval model with Dirichlet prior smoothing~\citep{zhai_study_2004} with $\mu = 2500$, which we will refer to as the \textbf{LM.Dir} model.
Both baselines are useful for comparison purposes.
There are two temporal ranking baselines: \textbf{Recency}~\cite{li_time_2003} and \textbf{KDE(score)}~\cite{efron_temporal_2014}, a state-of-the-art temporal feedback method.
Finally, the strongest baseline \textbf{LTR} is a learning to rank model with the full set of non-temporal features listed in \Cref{tab:non-temporal-features}.

\subsection{Results}
Using the described setup we performed an evaluation of these temporal ranking methods and present the results in~\Cref{tab:temprank}.
All methods evaluated use no future evidence, relying only on information that is pre-indexed and that can be processed at query time.

\textbf{Retrieval performance over all queries.} First, we compare the performance of standard text retrieval baselines over the full set of queries (All).
The IDF ranking function provides higher performance than the others in MAP and P30.
The intuition for this result is that term-frequency is not very important for ranking short texts, and therefore both BM25 and LM.Dir can hurt the retrieval performance by over-weighting term-frequency.
This also surfaces more spam documents with repeated words for these models.

Second, both temporal ranking baselines outperform text retrieval baselines.
The KDE(score) method outperforms Recency with a MAP result of 0.4455 (Recency with 0.4297).
These two baselines use only the collection as a single source of temporal relevance evidence.

Finally, learning to rank with multiple temporal features delivered the best results.
The proposed method RMTS, which combines temporal evidences of the collection with several additional sources of temporal signals from the Web, obtains the best results in MAP as well as P30 outperforming the state-of-the-art methods. 
We annotated \Cref{tab:temprank} with symbols denoting the statistically significance level ($p < 0.01$ or $p < 0.05$) of differences in effectiveness with respect to the stronger baselines: LTR and KDE(score) methods.

RMTS produced statistically significant improvements for MAP and P30. MAP improved 0.0281 over LTR using non-temporal features. This corresponds to a relative improvement of 6.2\%.
We follow the recommendations of~\citet{sakai_statistical_2014} to report the results and statistical significance tests.
According to a two-sided paired t-test for the difference in MAP $\mean{d} = 0.0281$ (with the unbiased estimate of the population variance $V = 0.0020$), RMTS statistically significantly outperforms the LTR model $(t(54) = 4.67$, $p < 0.000020$, $ES_{pairedt} = 0.64$, 95\% CI $[0.0161,0.0400]$).
Additionally, according to a two-sided paired t-test for the difference in MAP $\mean{d} = 0.0355$ (with the unbiased estimate of the population variance $V = 0.0049$), RMTS statistically significantly outperforms KDE(score) $(t(54) = 3.73$, $p < 0.000468$, $ES_{pairedt} = 0.51$, 95\% CI $[0.0165,0.0544]$).

Most surprisingly, P30 achieved a very competitive result that is in the same range as the top runs of TREC Microblog 2014.
Considering that the top TREC systems use more elaborate techniques such as lexical pseudo-relevance feedback (PRF) and Ensemble methods, this result becomes an important takeaway message: temporal signals from the crowds provide key information for microblog search.
According to a two-sided paired t-test for the difference in P30 means $\mean{d} = 0.0236$ (with the unbiased estimate of the population variance $V = 0.0049$), RMTS statistically significantly outperforms LTR $(t(54) = 2.48$, $p < 0.0164$, $ES_{pairedt} = 0.34$, 95\% CI $[0.0047,0.0426]$).
Additionally, according to a two-sided paired t-test for the difference in P30 means $\mean{d} = 0.0430$ (with the unbiased estimate of the population variance $V = 0.0117$), RMTS statistically significantly outperforms KDE(score) $(t(54) = 2.92$, $p < 0.005126$, $ES_{pairedt} = 0.40$, 95\% CI $[0.0137,0.0723]$).

\textbf{Temporal Query Performance Prediction.} TREC query topics consist of both temporal and nontemporal queries, therefore it could be advantageous to classify the query as to selectively use the temporal crowd signals or not.
Automatically using temporal features for ranking can lead to performance degradation and lower retrieval effectiveness if the query is non-temporal. Despite the low number of training queries (only 60) we discriminate the TREC 2013 queries into two classes using a simple strategy: we trained a temporal model with the 32 queries that have news articles and a second non-temporal model with 28 queries that do not have news articles. TREC 2014 queries were used for testing and were split into 38 temporal queries and 17 atemporal queries -- results are shown in \Cref{tab:temprank}.

Results showed that this improved results over LTR (T+A column in  \Cref{tab:temprank}): the MAP result of 0.4829 was the best result in all experiments. According to a two-sided paired t-test for the difference in means $\mean{d} = 0.0258$ (with the unbiased estimate of the population variance $V = 0.0038$), RMTS(T+A) statistically significantly outperforms LTR(T+A) $(t(54) = 3.09$, $p < 0.0031$, $ES_{pairedt} = 0.42$, 95\% CI $[0.0092,0.0424]$).  

In the atemporal set of queries, Recency~\cite{li_time_2003} and KDE(score) \cite{efron_temporal_2014} proved to be more vulnerable to the temporal evidence when the query is not temporal. This is probably due to the fact that both models always assume the query to be temporal, relying exclusively on the single source of temporal evidence that they have. Thus, query temporal classification in microblog search is a promising, yet difficult, research direction that can further improve time-aware ranking models.

\begin{figure*}[!htbp]
\centering
    \includegraphics[trim=0 5 0 5, clip, width=0.95\textwidth]{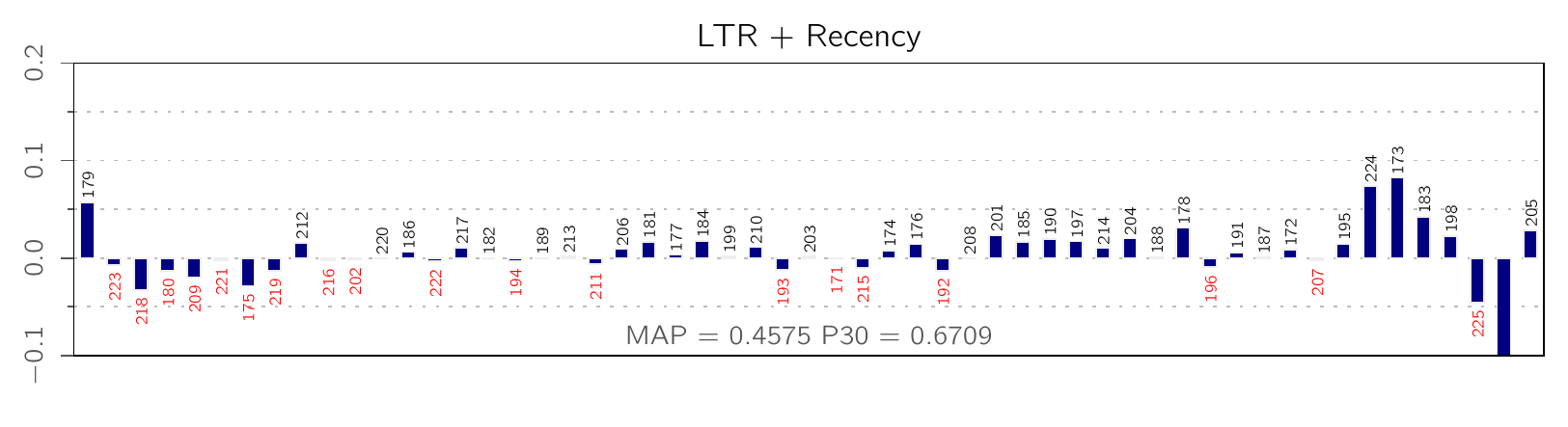} \\
    \includegraphics[trim=0 5 0 5, clip, width=0.95\textwidth]{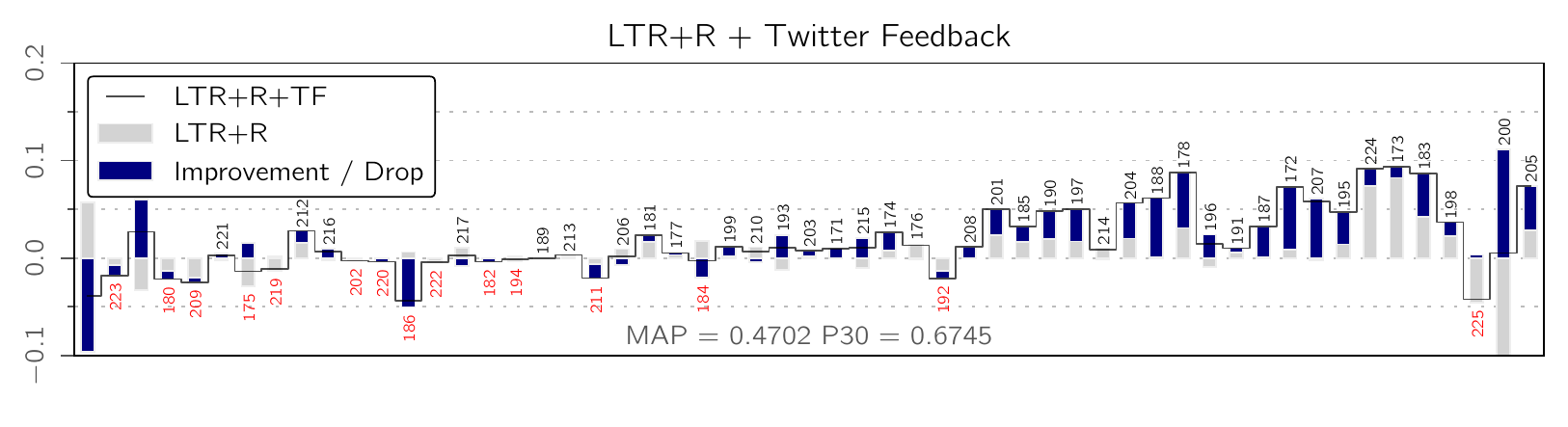} \\
    \includegraphics[trim=0 5 0 5, clip, width=0.95\textwidth]{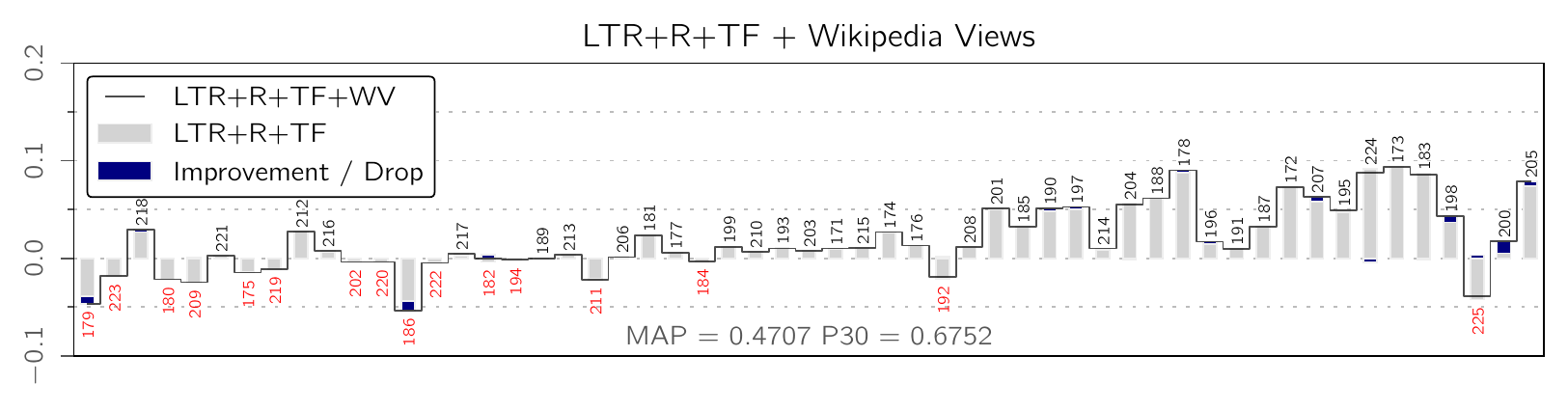} \\
    \includegraphics[trim=0 5 0 5, clip, width=0.95\textwidth]{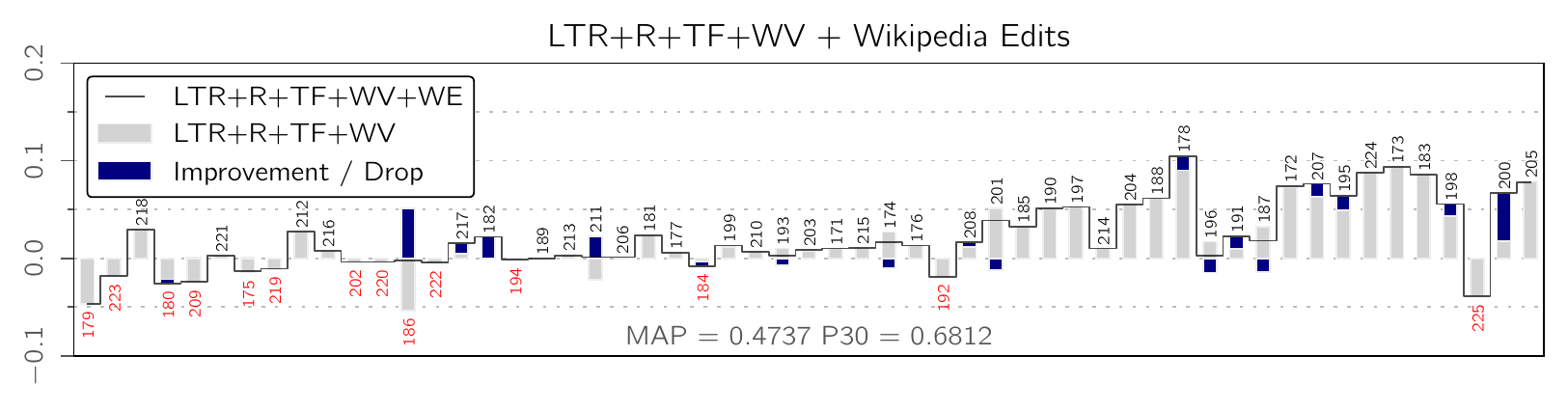} \\
    \includegraphics[trim=0 5 0 5, clip, width=0.95\textwidth]{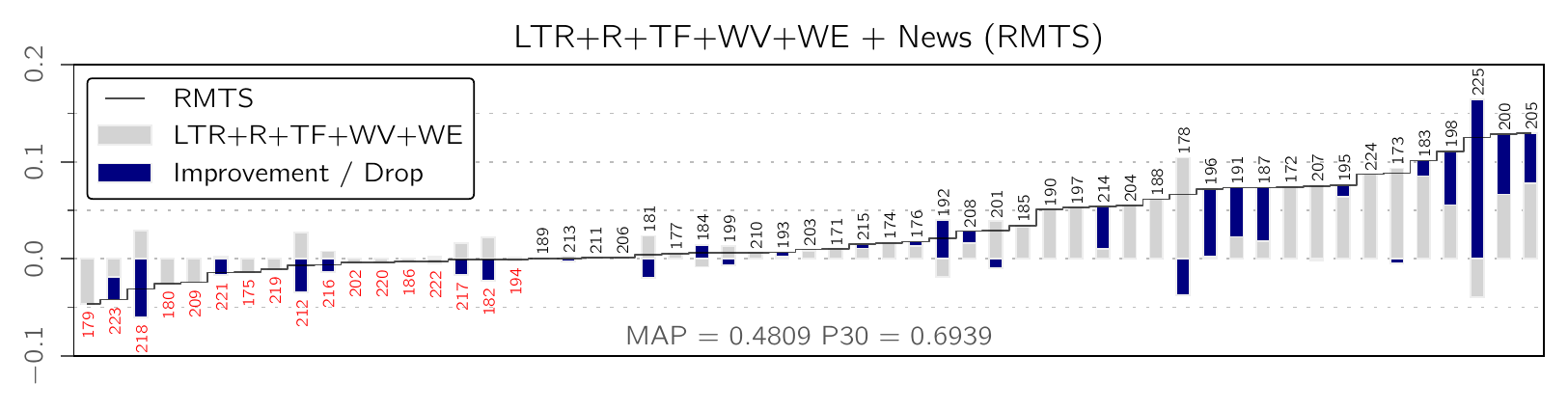} \\
    \caption{Per-feature retrieval results of the RMTS model: graphs show AP relative improvements over LTR by adding each temporal feature incrementally to the LTR model. Each graph illustrates the per-query results, where bars are labeled with the TREC topic number. Topic labels appear above/below if the performance improved/dropped relative to LTR.}
\label{fig:per-query-diffs-ltr}
\end{figure*}

\section{Discussion} 
In this section we provide a discussion of the results in three parts.
First, a per-feature and per-query analysis of the RMTS ranking model.
Second, we examine the contributions of the individual sources to the final ranking.
Finally, we examine the robustness of RMTS to missing sources.

\subsection{Per-feature and per-query analysis}
In general temporal ranking methods have a tendency to improve the overall performance of a system at the cost of degrading the performance of a number of other queries.
In~\Cref{fig:per-query-diffs-ltr} we present the per-query differences in Avg. Prec. relative to the LTR system to analyse the cumulative gains in performance of each feature across the set of test queries.
We found that the proposed method can indeed improve over the LTR in most queries (36 out of 55). Only a few queries lose performance when using multiple sources. This is an indication that some sources were incorrectly selected or provided low quality temporal evidence. Such effect has been widely recognized in temporal query performance prediction work. In these experiments, we decided not to filter sources, and instead use all sources for every query to best understand their role in the final retrieval model.

The use of \emph{recency priors} improves the Avg. Prec. of many queries that are more recency biased, however not all temporal queries favor recency, therefore the performance of those is degraded with this method.
It is clear that some queries are not temporal: the \emph{recency priors} and the temporal feedback models could not improve the rank of LTR in 29 and 26 queries respectively (from a total of 55 test queries).

Our model RMTS could not improve on 17 queries.
While the difference is on only {\raise.17ex\hbox{$\scriptstyle\sim$}}10\% of the total queries, our interpretation is that the queries are not all temporal. This is further confirmed by looking at the query topics.
Topic MB223 (\enquote{dog off leash}) for instance seems to be a non-temporal query that does not improve with temporal ranking.
In this case, every temporal source introduced noise and Avg. Prec. dropped almost 0.05.
Of note, the queries MB179 (\enquote{care of Iditarod dogs}) that improves with the recency feature and Topic MB212 (\enquote{Kate Middleton maternity wear}) that improves with Recency and Twitter Feedback, but their initial gains are inverted in the final model.

In the third graph it appears that the Wikipedia Views feature does not contribute much to ranking.
However, the reason is that the learned weight for this feature was found to be low in the complete RMTS model.
As seen in~\Cref{tab:temprank-sources} the Wikipedia Views feature improved results when used independently.
Moreover, it seems that withholding the WV feature improves the results slightly in terms of MAP but not P30 (see \Cref{tab:temprank-sources-removed}).

Finally, we also note that not all queries used the full range of external sources. For some queries it was not possible to find relevant news articles or select a Wikipedia article to provide useful evidence in the time-span of the corpus.

\subsection{Contributions of Individual Sources} 
The use of multiple temporal crowd signals calls for a deeper analysis of the contributions that each source makes to the final rank. In this experiment we examine the contribution of each temporal source to the improvement in MAP over the LTR model.
By looking at the results on \Cref{tab:temprank-sources}, we can observe that relative improvements in MAP with each individual temporal signal are in the range of +2.67\% to +4.51\%, reaching +6.02\% with RMTS. The temporal crowd signals that offer the most balanced MAP and P30 improvements are the News sources and Wikipedia edits.

The best improvement in MAP is obtained with the Twitter Feedback feature and the best improvement in P30 is obtained by the Wikipedia Edits feature. However, we observe that no source is particularly more effective than others across both retrieval metrics. This further evidences the stability of the proposed method.
In summary, all $f_{s_k}$ signals have a positive impact in the scores, but when combined as a whole, they outperform all other models with a MAP of 0.4809 and P30 of 0.6939. Actually, this is in line with our hypothesis: each temporal signal $f_{s_k}$ can provide different relevant time periods.
\begin{table}
	\centering
	\begin{tabular}{@{}p{2cm}p{2cm}p{2cm}@{}}
		\toprule
		Method        & MAP         & P30    \\ \midrule
		LTR           & 0.4528      & 0.6703 \\ \midrule
		LTR$_{+R}$    & +2.67\%\wl  & +1.36\% \\
		LTR$_{+TF}$   & +4.51\%\bl  & +0.54\% \\
		LTR$_{+WV}$   & +2.67\%\wl  & +1.00\% \\
		LTR$_{+WE}$   & +2.78\%\wl  & +2.16\% \\
		LTR$_{+News}$ & +3.05\%\bl  & +1.72\% \\ \midrule
		RMTS          & +6.21\%\bl & +3.52\%\wl \\ \bottomrule
	\end{tabular}
	\caption{Contributions of individual temporal sources to the ranking. Figures are relative improvement over non-temporal baseline. Symbol {\textnormal\wl} stands for a $p < 0.05$ statistical significant improvement over LTR (\textnormal{\bl} for $p < 0.01$).}
	\label{tab:temprank-sources}
\end{table}

\begin{table}
	\centering
	\begin{tabular}{@{}p{2cm}p{2cm}p{2cm}@{}}
		\toprule
		Method            & MAP           & P30    \\ \midrule
		RMTS              & 0.4809        & 0.6939 \\ \midrule
		RMTS$_{-R}$       & $-$0.94\%\wl  & $-$1.66\%\wl \\
		RMTS$_{-TF}$      & $-$1.31\%     & $-$1.12\% \\
		RMTS$_{-WV}$      & +0.48\%       & $-$0.95\% \\
		RMTS$_{-WE}$      & $-$1.10\%\wl  & $-$0.76\% \\
		RMTS$_{-News}$    & $-$2.74\%\wl  & $-$1.90\%\wl \\ \bottomrule
    \end{tabular}
	\caption{Ranking robustness to missing sources. Symbol \textnormal{\wl} stands for a $p < 0.05$ statistical significant difference compared to the RMTS model with all the temporal features (\textnormal{\bl} for $p < 0.01$).}
	\label{tab:temprank-sources-removed}
\end{table}

\subsection{Robustness to Missing Sources} 
\label{sec:robustness}
To further examine the robustness of the proposed method we studied the impact that a missing temporal source would cause in the final rank. This is a quite practical aspect in a real production system, where a temporal source might be temporarily unavailable. \Cref{tab:temprank-sources-removed} presents the MAP and P30 results.
There are two key facts arising from this table: the first one is related to the influence of the News temporal signals, and the second one concerns the low drop in MAP performance caused by missing any other temporal source.
In general MAP results are only slightly affected by each individual missing source -- in the worst case, the MAP drops 2.74\% when withholding the News. We consider this to be an excellent measure of the robustness to faulty temporal sources. Moreover, this further hints that multiple sources can provide complementary temporal signals.
On the one hand, the most important fact we take from these results is that the News is a key source of temporal signals.
On the other hand, the News feature is a group of news sources and not a single news source (e.g. AP).
Thus, the drop in the number of temporal information with its removal could be considered greater compared with the other features.

\section{Conclusions and Future Work}
\label{sec:conclusion}
This paper proposed the RMTS framework that mines the behavior of the crowds for temporal signals.
This new time-aware ranking method integrates lexical, domain and temporal evidences from multiple Web sources to rank microblog posts.
It explores the signals from: Wikipedia (through page views and page edit history), news articles, and Twitter feedback to estimate the temporal relevance of search topics.

\textbf{Retrieval precision.} We evaluated our system using the experimental setup used in the evaluation campaigns for microblog search at TREC 2013 and TREC 2014.
Experiments confirmed our hypothesis: the proposed approach offered a relative improvement of 13.2\% over BM25 and the Language Model with Dirichlet smoothing and 6.2\% over a strong learning to rank baseline with several lexical and domain features. Both improvements were statistically significant.

\textbf{RMTS is less biased.} A key advantage of the proposed RMTS model is its robustness and stability: the improvement over the LTR model (non-temporal features) could not be pin-pointed to a single source of temporal evidence, moreover, the retrieval model is tolerant to faulty temporal sources.

\textbf{Unified representation of temporal signals.} The proposed framework offers a principled methodology for mining and representing temporal signals from multiple crowds. It allows predicting temporal relevance from heterogeneous pairs of timestamps and weights mined from very diverse sources.

\textbf{Effective use of Wikipedia temporal signals.} The behavior dynamics of Wikipedia users is an effective source of temporal evidence for time-aware ranking. 
Previous works exploit article views statistics \cite{ciglan_wikipop_2010} and edit history \cite{georgescu_extracting_2013, steiner_mj_2013} for detecting events and entities related to the events, however to the best of our knowledge this is the first work that explores the use of multiple external sources for time-aware ranking. 

There are two main challenges that we wish to further explore.
News sources are a relatively easy resource to crawl, therefore the next steps could delve into the scaling of the number of news sources crawled to obtain better coverage of topics and other advantages.
However, new research questions arise: for a query $q$ \emph{which news sources should be selected?} and, \emph{how to weight them for each query?}
The second challenge is related to the query performance prediction. Temporal query classification is a difficult task as reported by a recent work by \citet{kanhabua_learning_2015}, where they studied the problem of detecting event-related queries in Web search streams (query logs). This is the second open problem that we will address as future work.

\vspace{2mm}
\textbf{Acknowledgments.} This research was partially supported by the EU H2020 project COGNITUS Grant 687605.
%
\bibliographystyle{abbrvnat-noplaces}
{
\bibliography{2016-wsdm-barbara-made-the-news}
}

%
%

\end{document}